\documentclass[twocolumn,showpacs,preprintnumbers,amsmath,amssymb]{revtex4}

\usepackage{graphicx}
\usepackage{dcolumn}
\usepackage{bm}

\begin{document}
\title{Effects of spin current on ferromagnets \\
}

\author{Z. Li, J. He and S. Zhang}
\affiliation{ Department of Physics and Astronomy, University of
Missouri-Columbia, Columbia, MO 65211 }

\date{\today}

\begin{abstract}
When a spin-polarized current flows through a ferromagnet, the local
magnetization receives a spin torque. Two consequences of this spin
torque are studied. First, the uniformly magnetized ferromagnet
becomes unstable if a sufficiently large current is applied. The
characteristics of the instability include spin wave generation
and magnetization chaos. Second, the spin torque has
profound effects on the structure and dynamics of the magnetic
domain wall. A detail analysis on the domain wall mass, kinetic
energy and wall depinning threshold is given.
\end{abstract}


\maketitle
\section{introduction}
There are many interesting phenomena generated by electronic
current. In this paper, we focus our discussions on two effects: domain wall
dynamics and magnetization instability. Recently, both topics have
received considerably interests in experiments \cite{Gan,
Koo,Fert,Parkin,Cowburn,Shinjo,Vaz2,Vila,Tsoi,Chien,Kent} and
in theories \cite{Li0,Stile2,Li1,Bazaliy, Waintal0, Tatara, Zhang, Shibata}.
It is shown that the current is able to displace
magnetic domain walls in a spin valve \cite{Fert}, in a constricted
nanowire \cite{Parkin}, in U-shaped (or L-shaped) nanowires
\cite{Cowburn,Shinjo}, in a ring structure \cite{Vaz2} and in zigzag
wires \cite{Vila}.  The current also generates dynamic domain walls
(time-dependent wall structure) via the generation of spin waves in
the uniformly magnetized ferromagnet. The point-contact experiments
on a single layer ferromagnet \cite{Tsoi,Chien,Kent} are likely
associated with the interface spin waves excitations \cite{Li0}.

The physics of these experimental results is believed to be
the consequence of spin angular momentum transfer, originally
proposed by Slonczewski \cite{Slon} and Berger \cite{Berger} in
magnetic multilayers. The extension of Slonczewski's model to a
homogeneous ferromagnet has been carried out by several groups
\cite{Bazaliy,Waintal0,Tatara,Li1}. Recently, we introduce two spin
torques in ferromagnets \cite{Zhang}. With our spin torque
model, most of the experimental observations can be quantitatively
analyzed. This paper is organized as follows. In Sec.~II, we briefly
review the mechanism of the spin torque in a single ferromagnet. In
Sec.~III, we study the bulk and surface spin waves by applying the spin
torque model. Several key characteristics of
the magnetization instability driven by spin torques are predicted.
In Sec. IV, we analyze the domain wall dynamics with and without
defects. We also construct an
analytic 1-D model to help understanding the key features in
simulations. Finally, an outlook for the effects of spin torques in
ferromagnets is presented in Sec. V.

\section{mechanism of spin torques in ferromagnets}

The spin-polarized current does not interact with spatially uniform
magnetization, except the classical Zeeman coupling between the
current-induced magnetic field and the magnetization.
In a real ferromagnet, the magnetization is not
spatially uniform due to the presence of various magnetic
interactions, which tend to break the ferromagnet into domains. Even
in a small particle, where the single domain assumption is
approximately valid, the thermal fluctuation leads to a
time-dependent non-uniform magnetization. In response to the
spatially non-uniform magnetization, the spin
current will be position-dependent and the spatial varying spin current
produces the mechanism
of the spin angular momentum transfer. To determine the spin
transfer torque, one requires simultaneously solving the
magnetization dynamics and the time-dependent spin current, which
would be a very complex problem. Fortunately, the dynamics of the
magnetization is much slower than that of the transporting
electrons, then it is a good approximation to work on the transport
equation with magnetization "frozen" spatially and temporally.
Similar to the calculation of electron dynamics in an atom: the electronic
structure is calculated by freezing the dynamics of nuclei.

Most of the theories of spin torques are built on the
above approximation. Additional assumptions are needed
to derive analytical expressions of the spin torque.
Berger \cite{Berger} first
investigated the current-induced domain wall motion and introduced
the ``domain drag force'' by using an intuitively physics picture
that the current can drag the domain wall moving along the path of
the current flow via an {\em s-d} exchange interaction. Since the
spin torque was not mathematically formulated in this work, it is
unclear how to solve the problem of domain wall dynamics for a
realistic magnetic wire by using this intuitive approach. Bazaliy {\em et al}
\cite{Bazaliy} proposed a spin torque model in a ferromagnet within the
ballistic transport model for half-metallic materials and they found
that the spin torque is ${\mbox{\bf
$\tau$}}$ $ \propto ({\bf j}_{e} \cdot {\bf \nabla \hat{\bf M}})$
where ${\bf j}_{e}$ is the electric current. The essential assumption
is that the spin polarization of the current is parallel to the
local magnetization, i.e., an adiabatic approximation. The above explicit
expression of the spin torque can be immediately combined with the
well-known LLG equation to calculate the response of magnetization
to the spin torque. Waintal and Viret \cite{Waintal0} has extended
this approximation by relaxing the adiabatic approximation.
They have shown that
the spin polarization of the current
is not parallel to the local magnetization in
a ballistic transport model. However, no clear mathematical
expression is given to implement this additional spin torque to the LLG
equation. Tatara and Kohno \cite{Tatara} proposed two spin current
effects: an adiabatic torque mentioned above and a momentum transfer
torque. The momentum transfer effect originates from the momentum
scattering by a domain wall and it is proportional to wall
resistivity. This momentum transfer effect is negligible except for
very thin walls.

We have proposed the spin torque by evaluating
the response of the conduction electron spins in a spatially and
temporally varying magnetization $\bf M({\bf r},{\it t})$ in the
semiclassical transport theory \cite{Zhang}. The spin torque has
been formulated in the following form
\begin{equation}
{\bf \tau}_s = b_{J}(\hat{\bf j}_e \cdot \mbox{\boldmath $\nabla$}
) {\bf M}-\frac{c_{J}}{M_s}{\bf M} \times (\hat{\bf j}_e \cdot
\mbox{\boldmath $\nabla$} ) {\bf M}
\end{equation}
where ${\hat{\bf j}}_e$ is the unit vector in the direction of the
current flow, $b_J= Pj_e \mu_B/eM_s$, $P$ is the spin polarization
of the current, $M_s$ is the saturation magnetization, $\mu_B$ is
Bohr magneton, $c_J = \zeta b_J$, and $\zeta$ is a dimensionless
constant that describing the degree of the nonadiabaticity between
the spin of conduction electrons and the local magnetization. For a
typical ferromagnet (Ni,Co,Fe and their alloys), $\zeta$ is within a
range of $0.001\sim0.05$. The physical interpretation is that the
$b_J$ term is an adiabatic spin torque, describing the adiabatic
process of the non-equilibrium conduction electrons, i.e., the
direction of the spin polarization of current is parallel to
the local magnetization within a domain wall; the $c_J$ term is a
non-adiabatic torque, which is related to the spatial mistracking of
spins between conduction electrons and local magnetization.

The advantage of expressing the spin torques in the form Eq.~(1)
is that one can readily generalize LLG equation in the presence of
the spin current,
\begin{equation}
\label{LLG} \frac{\partial {\bf M}}{\partial t}=-\gamma {\bf
M}\times {\bf H}_{eff}+\frac{\alpha}{M_s}{\bf M} \times
\frac{\partial {\bf M}}{\partial t} + {\bf \tau}_s
\end{equation}
where $\gamma$ is the gyromagnetic ratio, and ${\bf H}_{eff}$ is the
effective magnetic field including the external field, the
anisotropy field, magnetostatic field, and the exchange field, and
$\alpha$ is the Gilbert damping parameter. The current-driven
magnetization dynamics will be studied by solving this generalized
LLG equation in various situations.

\section{Spin wave excitations}
In this section, we apply our generalized LLG equation, Eq.~(2) to
study the spin wave excitations in a single layer magnetic film. The
local effective field is
\begin{equation}
{\bf H}_{eff} = \frac{H_K M_x}{M_s} {\bf e}_x + \frac{2A}{M_s^2}
\nabla^2 {\bf M} - 4\pi M_{z} {\bf e}_z + H_{e} {\bf e}_x
\end{equation}
where $H_K$ is the anisotropy field, alone the $x$-axis, $A$ is the
exchange constant, and we include a self-demagnetization field $4
\pi M_z$ of the film. The initial magnetization saturates in the
direction of the magnetic field $H_e$ along the $x$-axis in the
plane of the layer.

When one applies a sufficiently large current along the $x$-axis,
the uniformly magnetized film becomes unstable. To see this, we
consider a small deviation of the magnetization vector from the easy
axis ${\bf e}_x$,
\begin{equation}
{\bf M}=M_s {\bf e}_x+ \delta {\bf m}e^{i(\omega t+ {\bf k}\cdot
{\bf r})}
\end{equation}
where $\delta {\bf m}$ is a small vector. Inserting Eqs.~(3) and
(4) into Eq.~(2), and keeping only the terms linear in $\delta
{\bf m}$, we obtain two linearized equations for $\delta m_{y}$
and $\delta m_{z}$. A secular equation is then established for the
spin wave frequency $\omega$ and the spin wavevector ${\bf k}$.
\begin{equation}
{\left[\begin{array}{cc}-i(\omega-b_{J}k_{x}) &
A_{1} \\
A_{2} & -i(\omega-b_{J}k_{x})
\end{array}\right]\left[\begin{array}{cc}\delta m_{y}
\\\delta m_{z}\end{array}\right]=0}
\end{equation}
where we have defined $A_{1}=-\gamma\left(\frac{2A}{M_s}k^{2}+4\pi
M_s+H\right)-i\alpha\omega+ic_{J}k_{x}$ and
$A_{2}=\gamma\left(\frac{2A}{M_s}k^{2}+H\right)+i\alpha\omega-ic_{J}k_{x}$,
in which $k^2 =k_x^2+k_y^2+k_z^2$ and $H=H_{e}+H_{K}$.

The above linearized equations have a non-zero solution if and
only if
\begin{equation}
{\rm det } \left[\begin{array}{cc}-i(\omega-b_{J}k_{x}) &
A_{1} \\
A_{2} & -i(\omega-b_{J}k_{x})
\end{array}\right] =0
\end{equation}
The above equation establishes the relation between the spin wave
frequency $\omega$ and the wavevector ${\bf k}$. For a real
wavevector ${\bf k}$, $\omega$ is a complex number. If the
imaginary part of $\omega$ becomes negative, Eq.~(4) is
exponentially growing with time. In this case, the uniform
magnetization becomes unstable. One may define the critical
current such that $Im~\omega =0$. From Eq.~(6), we find $Re~\omega
=c_{J}k_{x}/\alpha$ and
\begin{eqnarray}
b_{J}-c_{J}/\alpha=\nonumber\\
\frac{\gamma}{k_x} \sqrt{ \left(\frac{2A}{M_s}{\it k}^{2}+H
\right)\left(\frac{2A}{M_s}{\it k}^{2} +H +4\pi M_s \right)}
\end{eqnarray}

Equations (7) gives out the current density required to generate
spin waves with a given wave vector ${\bf k}$. Interestingly, the
minimum current density does not occur at the uniform mode of ${\bf
k}=0$ as in the ordinary spin wave excitations, rather the spin wave
with a finite wavevector is first excited by the current. To see this, we
minimize $b_J-c_J/\alpha$ in Eq.~(7) with respect to the
wavevector, and we find the minimum current density occurs at
\begin{equation}
k_x^2 = \frac{M_{s}}{2A} \sqrt{(4\pi M_s +H)H}.
\end{equation}
and $k_y=k_z=0$. The corresponding wavelength is,
\begin{equation}
\lambda_c =\frac{1}{k_x}=\sqrt{\frac{2A}{M_{s}}} \left[ (4\pi M_s
+H)H \right]^{-1/4}.
\end{equation}

The new length scale given above places a severe limitation on
micromagnetics: one needs to choose mesh size to be smaller than
$\lambda_c$ when the current density exceeds the critical current
density, in order to correctly capture the current-driven effects in
the simulation. For certain parameters, particularly at large
magnetic fields, $\lambda_c$ can be comparable or smaller than the
exchange length.

Inserting Eq.~(8) back to Eq.~(7), we find that
the minimum current density for the instability is,
\begin{eqnarray}
|b_J-c_J/\alpha|_{min} = \gamma\sqrt{\frac{2A}{M_s}} \left(
\sqrt{H}+\sqrt{H+4\pi M_s} \right) .
\end{eqnarray}

Note that we have assumed a spin wave spectrum in the form of
Eq.~(8) which represents a spin wave in an isotropic infinite
medium. If we take $\zeta=\alpha$ the spin wave instability does not
occur for any large spin currents.

In experiments, the current may not uniform across the sample. For example,
the current density is very large at the contact area in the
point-contact experiments and it becomes negligibly small away from the
contact area. In this case, the excited spin waves
are confined in a narrow region near the interface. To access the
spin wave instability in this situation, we consider a surface spin
wave mode
\begin{equation}
{\bf M} = M_s {\bf e}_x + \delta  {\bf m} e^{-x/\kappa}
e^{i(\omega t + k_{y} y + k_{z} z ) } .
\end{equation}
where $\kappa$ is the penetration length.

By inserting it into Eq.~(2) and by repeating the derivation
similarly, we find the instability $Im~\omega = 0$ occurs at
\begin{equation}
-(b_{J}+\alpha c_J)
=\gamma\alpha\kappa\left(\frac{2A}{M_s}(k_{y}^{2}+k_{z}^{2})+H
+2\pi M_s\right).
\end{equation}

Note a negative sign on the left-hand side; it indicates that the
surface spin waves can only be generated by the current flowing
along one direction. This result is consistent with the
point-contacted experiments \cite{Chien}, in which the current
flowing from the magnetic layer to the nonmagnetic tip is able to
excite spin waves. The magnitude of the critical current for
surface spin wave is proportional to the damping parameter
$\alpha$ and the penetration length $\kappa$. We note that the
exchange field at the surface for the localized spin wave is
$(2Aa_0/M_s^2) \nabla_x {\bf M}$ where $a_0$ is the lattice
constant so that the exchange field inside the layer
$(2A/M_s^2)\nabla_x^2 {\bf M}$ will be compensated. This is the
reason that the right hand side of Eq.~(12) does not contain the
$k_x^2$ term.

An estimation on the magnitude of spin wave instability can be
readily done by using the materials parameters of Co:
$\gamma=1.9\times10^{7}~(Oe)^{-1}s^{-1}$, $4\pi M_s
=1.8\times10^{4}~Oe$, $H_K =500~Oe$, $M_s=14.46\times10^{5}~ A/m$,
$A=2.0\times10^{-11}~ J/m$ and $P=0.35$. If one takes the damping
parameter to be 0.01, the non-adiabaticity $\zeta=0.02$, the
penetration length of $5~nm$, and $k_y=k_z \approx 0$ for the long
wave length limit, we find that the critical current at $H_e=0~Oe$
is $j_{min}^{bulk}= 1.12\times 10^{10}~A/cm^2$ for the bulk spin
wave and $j_{min}^{surface}=6.4\times10^{7}~A/cm^{2}$ for the
surface mode, which is two orders of magnitude smaller than that
of bulk spin wave.

The fact that the large enough current density can excite spin waves
with different wave lengths raises a question: what is the
magnetization state at the large current density? Shibata {\em et
al} \cite{Shibata} predicted that a {\em static} multi-domain state
can be formed if the current exceeds a second critical current
higher than the critical current density defined above. By using
Eq.~(16), we find that the {\em static} domain configuration is
unlikely to form. This is because the domain will be moving with an
average velocity determined by the non-adiabatic torque $c_J/\alpha$
\cite{Miltat0,Zhang} seen in next section. We had reported earlier
that a large current density drives the uniform magnetization into
spatially and temporally chaotic motion \cite{Li3}. The detail
analysis on this transition will be given elsewhere.

\section{Domain wall motion}

One of the most interesting predictions on the domain wall motion is
that a steady domain wall velocity is independent of $b_J$. Rather
the steady state velocity is solely determined by the non-adiabatic
torque and the external magnetic field. We show the proof below.

In a steady state motion, one can assume the magnetization vector
${\bf M}={\bf M}(x-v_x t,y,z)$. Then, $\partial {\bf M}/\partial t
= -v_x \partial {\bf M}/\partial x $. For a current applied in
x-axis, the LLG equation, Eq.~(2), is thus
\begin{equation}
(b_J+v_x)\frac{\partial {\bf M}}{\partial x} - \frac{\alpha
v_x+c_J}{M_s} {\bf M}\times \frac{\partial {\bf M}}{\partial x} =
\gamma {\bf M}\times {\bf H}_{eff}
\end{equation}
By performing inner-product of ${\bf M}\times {\partial {\bf M}}/
{\partial x}$ with the above equation, one has
\begin{equation}
\frac{\alpha v_x+c_J}{M_s} \left| {\bf M}\times \frac{\partial
{\bf M}}{\partial x} \right| ^2 = \gamma {\bf H}_{eff} \cdot
\frac{\partial {\bf M}}{\partial x}
\end{equation}
We now integrate the above equation over the entire domain wall.
We find
\begin{equation}
\frac{\alpha v_x+c_J}{\gamma M_s}  \int dV \left|\frac{\partial
{\bf M}}{\partial x} \right|^2 =  \int dV \frac{\partial
E}{\partial x} = \frac{1}{v_x} \frac{\partial}{
\partial t} \int E dV
\end{equation}
where $E$ is the energy density that is defined $dE = -{\bf H}_{eff}
\cdot d{\bf M}$, and we have also used the fact that ${\bf M}$ and
$\partial {\bf M}/\partial x $ is perpendicular. The right-hand side of
the equation becomes $\pm 2S_0M_s H_e $ because the rate of the
total energy change for a uniformly moving wall is $2S_{0}M_{s} H_e
v_x $, where $S_0$ is the cross section perpendicular to the moving
wall. Therefore, the velocity is
\begin{equation}
v_x = - \frac{c_J}{\alpha} \pm \frac{\gamma H_e W}{\alpha}
\end{equation}
where we have defined the wall width $W$
\begin{equation}
W^{-1} = \frac{1}{2S_{0}M^{2}_{s}} \int dV \left|\frac{\partial
{\bf M}}{\partial x} \right|^2
\end{equation}

The striking conclusion from Eq.~(16) is that the wall velocity in
the absence of the magnetic field is exactly $-c_J/\alpha$ for any
type of walls as long as the wall is moving at a constant
velocity. Without the current, Eq.~(16) becomes the well-known
Walker's velocity \cite{Walker}. We should point out that the wall
width, Eq.~(17), is weakly dependent on the magnetic field and the
adiabatic torque $b_J$ \cite{Li1}, and thus the terminal wall
velocity will slightly depend on $b_J$ in the presence of the
field.

The wall velocity given by Eq.~(16) breaks down when the wall is
{\em not} moving with a constant velocity. To describe a non-uniform wall
motion, it is necessary to postulate an approximate wall structure.
Here, we follow Walker's procedure \cite{Walker} by introducing two
polar angles $\theta $ and $\varphi$ in the form mimic a transverse domain
wall \cite{footnotes}. By placing the explicit form of $\theta $ into
Eq.~(2), we find
\begin{eqnarray}
\frac{d\varphi}{d{\it t}}=\gamma (H_{e}-4\pi\alpha
M_{s}\sin\varphi\cos\varphi)+ \frac{\alpha
b_J-c_J}{W} \\
\alpha {\it v_{x}}({\it t})+ W\frac{d \varphi}{d{\it t}}=\gamma
H_{e}W-c_{J}
\end{eqnarray}
where $\varphi$ is the out-of-plane angle of the magnetization
vector, $v_{x}(t)$ is the velocity, and we have discarded
$\alpha^2$ terms.

To further gain insights of the wall velocity, we assume that the
distortion of the wall is small during
the wall motion. Then, we treat the wall width $W$ as
time-independent constant, and replace $\sin\varphi\cos\varphi$ by
$\varphi$ in Eq.~(18). By differentiating Eqs.~(18) and (19) with
respect to ${\it t}$, and by eliminating $\varphi$ in the resulting
equations, we find the equation for the wall velocity
\begin{equation}
m^{*}\frac{d {\it v_{x}(t)}}{d t}+\frac{2M_s\alpha}{\gamma W}
 {\it v_{x}(t)}-2M_{s}H_{e}+ \frac{2M_s}{\gamma W}c_J=0
\end{equation}
where we have defined the domain wall effective mass per unit area as
\cite{Doring},
\begin{equation}
m^{*}=(2\pi \gamma^{2} W)^{-1} .
\end{equation}
Equation (20) reveals that the domain wall can be treated as
a particle with an effective mass $m^*$, subjecting to a
friction force (second term) and external forces from the magnetic
field (third term) and the current (fourth term). Interestingly, the
adiabatic torque does not contribute to the current-driving force.
The solution of Eq.~(20) is
\begin{equation}
{\it v_{x}}({\it t})=-\frac{c_J-\gamma WH_e}{\alpha}+Ce^{-t/\tau}
\end{equation}
where $\tau=(4\pi M_{s}\gamma\alpha)^{-1}$ is the relaxation time
and $C$ is a constant determined by the initial condition. If one
assumes that the field or the current is applied at $t=0$, i.e.,
$\varphi (0) =0 $, we find, from Eq.~(18) and (19), $v_{x}(0) =
-b_J$. By using this initial value of the velocity, Eq.~(22)
becomes
\begin{equation}
v_{x}(t) = -\frac{c_J-\gamma WH_e}{\alpha} \left( 1 - e^{-t/\tau}
\right) -b_J e^{-t/\tau}
\end{equation}
One may understand the above two terms as follows. The first is a
``translational'' velocity and one might introduce a kinetic energy
\begin{equation}
E = \frac{1}{2} m^* v_{x}^2(t) =  \frac{(c_J-\gamma WH_e)^2}{4 \pi
\gamma^2 W \alpha^2} \left( 1 - e^{-t/\tau} \right)^2
\end{equation}
The time scale to establish a fully accelerated motion is
determined by $\tau$. For a Co wire, $\tau=0.3~ns$ if we take
$\alpha=0.01$. The second term of the Eq.~(23) is related to the
domain wall distortion, and therefore, it does not contribute the
mobility of the domain wall. The displacement of the domain wall
is $x_c = \int_{0}^{\tau}v_{x}(t) dt$.

We illustrate below that the concept of the ``translational'' kinetic energy
is helpful in understanding the domain wall motion in a nanowire
containing defects or pinning centers. For example, let us consider
a domain wall propagation from one place (A) to another place (B),
shown in Fig.~1(a). Between A and B regions, there is a distribution
of defects. We show that the minimum current-density or the minimum
magnetic field required to overcome these pinning centers depends on
the relative distance between the defects and the domain wall. If a
strong defect is located far away from the initial domain wall
position (A), the domain wall is able to develop its fully kinetic
energy, and the domain wall is able to pass through the pinning
center when the kinetic energy is larger than pinning potential. On
the other hand, if the defect locates near the initial wall center,
the kinetic energy is small when the domain wall encounters the
defect and thus a relative weak defect potential is able to trap the
domain wall. To be more quantitative, we perform the following
calculation.

We model a defect by introducing an artificial local anisotropy
pinning source in an otherwise perfect nanowire. The anisotropy of
the defect is $H_{d}=4H_{K}$ which is 4 times larger than the
anisotropy of the wire, $H_{K}=500~Oe$. We choose a
one-dimensional model, i.e., the magnetization does not vary in
the direction of width and thickness, to illustrate our points.
The easy axis of the wire and the defect, and the current or
magnetic field, are all along x-axis. The mesh size in x-direction
is $4~nm$. Before one turns on the current, a stationary N\'{e}el
wall is centered at $x=0$, and the defect is located $x_d$ from
the center of the wall. At $t=0$, a current is turned on and we
calculate the domain wall motion afterwards.

First, we have placed the defect far away from the region A, i.e.,
$x_d \gg \int_{0}^{\tau}v_{x}(t) dt$, we find that the critical
current is proportional to the damping parameter in the absence of
magnetic field, consistent with Eq.~(24). When we vary the
exchange constant which effectively changes the domain wall width,
we find that the critical current scales as $\sqrt{W}$, again,
consistent with Eq.~(24). We now vary the defect position. In
Fig.~1(b), we show the critical current as a function of the
defect position. As expected, the critical current decreases as
one places the defect away from the original wall. The initial
increase of the critical current at small distance comes from the
adiabatic torque, $b_J$. When the defect is very near the original
wall, the displacement of the wall by the $b_J$ term exceeds $x_d$
and thus the wall overcomes the defect potential by the adiabatic
torque. Therefore, at the small distance, the adiabatic torque is
more effective than the kinetic energy involved.
\begin{figure}
\centering
\includegraphics[width=6.5cm]{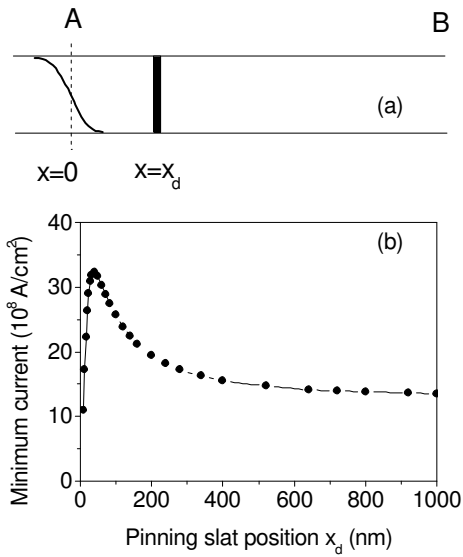}
\caption{(a) Schematic pinning source in a Co nanowire, (b) The
minimum spin current required for a wall propagation through a
pinning as a function of pinning position $x_{d}$. The parameters
of Co are $4\pi M_s =1.8\times10^{4}~{\it Oe}$, $H_K =500~{\it
Oe}$, $M_s=14.46\times10^{5}~A/m$, $A=2.0\times10^{-11}~J/m$,
$H_{e}=0$, $\zeta=0.02$ and $\alpha=0.006$}
\end{figure}

\section{outlook}
There are many fundamental issues on current-induced effects. We
intend to list a few of them below.

On the fundamental mechanism, the spin torque given by Eq.~(1) is
valid up to the first derivative in space. When the domain wall
thickness is extremely small in some nanoconstriction, the high
order terms become important. A better quantum mechanical
treatment for the spin torque is required to re-formulate the spin
torque in this case.

In a large field or a large current, the wall is usually moving
quite irregularly. Non-uniform wall motion appears: the wall can be
bounced back and forth in transverse directions, and the wall
changes its structure. For example, the current can drive vortex
wall motion relatively easier than the transverse wall in the
presence of defects. However, the vortex wall is usually unstable
during its motion and the vortex wall tends to change into a
transverse wall. These detailed complications have to be studied
via numerical solutions of the LLG equation \cite{Miltat0,He0}. To
model a realistic device, an extensive numerical effort is
required.

The effect of Joule heating by the current has not been
quantified. The experimental results show that the sample
temperature increases less than $5~K$ when the current density is
order of $10^{5}\sim 10^{6}~A/cm^{2}$ \cite{Fert}, while the
temperature is dramatically increased when the current density is
$10^{7}~A/cm^{2}$ \cite{Yamag}. These studies are for a steady
current density. In most of the devices, it is much desirable to
use a pulsed current source. A thorough study on the Joule heating
on the amplitude and duration of the current seems very important.
Frequently, the observed critical current is smaller than the
optimal theoretical value by a factor of two. Whether it is due to
heating is unclear at the present time.

The finite temperature theory on the domain wall motion is lacking.
It would be interesting to study the lifetime of a trapped domain wall and
it takes for the current to depinning the wall at finite temperature.
Similar to the effect of the finite temperature in current-driven
switching in spin valves, we expect that the role of temperature has
profound effects on domain wall motion.

The research was supported by NSF (ECS-0223568 and DMR-0314456).

\end{document}